\documentclass[sigconf]{acmart}

\setcopyright{none}
\renewcommand\footnotetextcopyrightpermission[1]{}

\usepackage{array}
\usepackage{multirow}

\begin{document}

\title{Case Survey Studies in Software Engineering Research}

\author{Jorge Melegati}
\email{jmelegatigoncalves@unibz.it}
\orcid{0000-0003-1303-4173}
\affiliation{%
  \institution{Free University of Bozen-Bolzano}
  \streetaddress{Piazza Domenicani 3}
  \city{Bolzano}
  \country{Italy}
  \postcode{39100}
}

\author{Xiaofeng Wang}
\email{xiaofeng.wang@unibz.it}
\orcid{0000-0001-8424-419X}
\affiliation{%
	\institution{Free University of Bozen-Bolzano}
	\streetaddress{Piazza Domenicani 3}
	\city{Bolzano}
	\country{Italy}
	\postcode{39100}
}


\begin{abstract}
\textit{Background:} Given the social aspects of Software Engineering (SE), in the last twenty years, researchers from the field started using research methods common in social sciences such as case study, ethnography, and grounded theory. More recently, case survey, another imported research method, has seen its increasing use in SE studies. It is based on existing case studies reported in the literature and intends to harness the generalizability of survey and the depth of case study. However, little is known on how case survey has been applied in SE research, let alone guidelines on how to employ it properly. \textit{Aims:} This article aims to provide a better understanding of how case survey has been applied in Software Engineering research. \textit{Method: } To address this knowledge gap, we performed a systematic mapping study and analyzed 12 Software Engineering studies that used the case survey method. \textit{Results:} Our findings show that these studies presented a heterogeneous understanding of the approach ranging from secondary studies to primary inquiries focused on a large number of instances of a research phenomenon. They have not applied the case survey method consistently as defined in the seminal methodological papers. \textit{Conclusions:} We conclude that a set of clearly defined guidelines are needed on how to use case survey in SE research, to ensure the quality of the studies employing this approach and to provide a set of clearly defined criteria to evaluate such work.\footnote{This is a postprint (accepted version) of: Jorge Melegati and Xiaofeng Wang. Case Survey Studies in Software Engineering Research. ACM / IEEE International Symposium on Empirical Software Engineering and Measurement (ESEM) (ESEM '20), October 8--9, 2020, Bari, Italy. \href{https://doi.org/10.1145/3382494.3410683}{\textcolor{blue}{https://doi.org/10.1145/3382494.3410683}}}
\end{abstract}



\pagestyle{plain}
\maketitle

\section{Introduction}
\label{sec:introduction}

In the last 20 years, Software Engineering researchers have broadened the research methods employed to perform their scientific studies. Since some aspects of software development regard human and organizational elements, researchers often employ methods imported from social sciences. The most evident approach is case study, employed by many studies in the Software Engineering literature. Other methods include ethnography (e.g.,~\cite{Passos2011,Boden2011}) and grounded theory (e.g.~\cite{Santos2014,Giardino2016}). Researchers, trained into so-called ``hard'' sciences such as computer science and software engineering, may encounter difficulties in employing research methods common in ``soft'' social sciences. To deal with that, several authors proposed guidelines specifically for Software Engineering researchers to employ these methods. Some examples are Runeson and H\"{o}st~\cite{Runeson2009} guidelines for case study, Passos et al.~\cite{Passos2012} for ethnography, and Stol et al.~\cite{Stol2016} for grounded theory.

Recently, some studies in the Software Engineering literature claimed to have used another method imported from other disciplines: case survey. It is an approach to identify and statistically test patterns across case studies~\cite{Larsson1993,Jurisch2013}. Its goal is to provide wider generalization than single or multiple-case studies based on rich data these numerous inquiries provide~\cite{Jurisch2013}. This method is valuable to Software Engineering research since it allows researchers to take advantage of an increasing number of case studies published and to improve the generalizability of the results, which is a common validity concern over case studies~\cite{Runeson2009}.

However, a quick search for examples from Software Engineering research indicated that case survey has been employed in various manners, ranging from studies with a large number of cases (e.g.,~\cite{Petersen2018}) to the analysis of a large number of online resources such as blog posts (e.g.,~\cite{Bajwa2017}). There is little consensus shown in these studies on what case survey is and how to employ it systematically in the Software Engineering research. Based on this observation, our goal is to assess how much the actual use of the case survey approach in the Software Engineering literature diverged from the seminal methodological guidance. Therefore, we proposed the following research question:

\begin{center}
	\textbf{RQ: How do Software Engineering researchers employ the case survey method?}
\end{center}

To achieve this goal, we performed a systematic mapping study on the Software Engineering literature to explore how the case survey method has been employed. We also compared the definition of and the concepts involved in the case survey method to those in the reviewed studies. Our results indicate that case survey has not been consistently applied in Software Engineering studies as defined in the seminal methodological papers.

The remaining of the paper is organized as follows: Section~\ref{sec:case_survey_method} presents a description of case survey as described in the seminal methodological papers since its initial proposal. In Section~\ref{sec:research_method}, we describe the systematic mapping study conducted and, in Section~\ref{sec:results}, its results. In Section~\ref{sec:discussion}, we discuss the results and Section~\ref{sec:conclusions} concludes the paper.

\section{The case survey method}
\label{sec:case_survey_method}

As the name suggests, the case survey method proposed to combine the advantages of two common research methods: case study and survey. Larsson~\cite{Larsson1993} wrote probably the most cited methodological paper on the method. In the article, he claimed that the first study to use the method was performed by Yin and Yates~\cite{Yin1974} in 1974. Nevertheless, in a paper from the same period~\cite{Yin1975}, the same group of researchers stated that an initial version of the method was previously employed in another study, published one year before, i.e.~\cite{Yin1973}. These researchers developed the method as a way to aggregate the existing research. They first employed the method on studies, mentioned above, focused on public policies. They aggregated the results of the case studies performed on different public programs to reach new insights and inform governmental decision-makers. 

From the same group, the first methodological paper, i.e.~\cite{Lucas1974}, emerged. In this early paper, the core aspect of the method was already present: ``qualitative and descriptive information found in case studies is put in a form susceptible to quantitative analysis''~\cite{Lucas1974}. The author highlighted the importance of a rigorous approach, divided into two areas: first, searching, selecting and sampling studies, and second, selecting and defining concepts across the studies and analyzing the results. Almost twenty years later, Larsson~\cite{Larsson1993} described the method in four steps:  i) select a group of existing case studies, ii) design a coding scheme to convert case descriptions into quantitative variables, iii) code the cases using multiple raters, and iv) analyze the results statistically. In summary, a case survey consists of a secondary study where primary case studies from the literature are compiled, and their data analyzed using quantitative techniques similar to those used in a survey. 

Larsson~\cite{Larsson1993} performed an analysis of how the case survey method had been used until that moment. To achieve the goal, Larsson divided further the four steps into 12 steps to compare the studies~\cite{Larsson1993}. These steps were:

\begin{itemize}
	\item Development of initial research questions: Larsson~\cite{Larsson1993} argued that case surveys could be used to test hypotheses or explore but stressed the importance of theory to select cases properly.
	\item Definition of case selection criteria: the criteria to select studies should be explicit and based on research questions. It should be based on how much data is reported rather than ``how, when, or where'' it was published. Besides that, a large number of cases is needed for statistical hypothesis testing.
	\item Collection of case studies: in this step, the search is executed. If needed, the researchers should add different sources or remove studies with not enough data. There should also be a concern if the sampled cases represent the population.
	\item Creation of the code scheme to convert the cases into variables: definition of the instrument that will guide the conversion of qualitative into quantitative data. It is determined by a trade-off between ``resource-saving, reliable simplicity, and information-rich complexity.''
	\item Coding the cases with multiple-raters: here, the raters use a defined scheme to code the cases. Several authors~\cite{Larsson1993,Jurisch2013} suggested that at least two, preferably three raters, are employed. 
	\item Participation of original authors: Larsson~\cite{Larsson1993} suggested that original case study authors could be employed as the third-rater for their studies. Since they had contact with primary data, they may have other insights since part of the data may have been left out of the report because of space constraints.
	\item Measure of inter-rater reliability: this aspect is a crucial measure of coding quality, but focusing too deep on it may be counter-productive~\cite{Larsson1993}.
	\item Resolution of coding discrepancies: here, several techniques could be employed to solve coding differences among raters.
	\item Statistically analysis of coding validity: evaluation if the chosen codes agree with the original author's results or other data available of the case.
	\item Statistical analysis of the impact of specific case characteristics: evaluation if case study aspects such as research design, publication, and time influenced the coding.
	\item Statistically analyzing the created case data set: using the quantitative data to draw conclusions.
	\item Reporting the results: develop the study report and, if applicable, publish it.
\end{itemize}

Another specific aspect of the method is the reader-analyst role~\cite{Yin1975}. Yin and Yates argued that these researchers are scientific observers, and their prime virtue ``is that they are trained observers and can make more difficult judgments than ordinary [survey] respondents''~\cite{Yin1975}. Larsson~\cite{Larsson1993} compared the coding schemes with questionnaires: ``the complex coding scheme allowed for richer data collection and analysis than a questionnaire survey could realistically be expected to achieve.''

In information systems literature, Jurisch et al.~\cite{Jurisch2013} proposed an adapted set of guidelines for the method, including the possible advantages of using it. The authors proposed five steps: developing a research question, searching and sampling case studies, design a coding scheme, transforming qualitative into quantitative data, and statistical analysis. Additionally, the authors focused on multiple raters. In Software Engineering literature, several guidelines for synthesizing empirical evidence mention case survey as a synthesis technique, e.g., ~\cite{Cruzes2011},~\cite{Guzman2014}, and ~\cite{Cruzes2015} but, to the best of our knowledge, no paper focused specifically on the method.

In summary, we can identify three key steps in definitions to the case survey method: selecting cases, converting qualitative into quantitative data, analyzing statistically the formed data set. Each step commonly consisted of some characteristics described below.

\textbf{Case selection} should gather the largest possible number of cases that were not necessarily \textbf{published} in peer-reviewed venues. Nevertheless, the \textbf{amount of available data} should be enough to describe the case to allow raters to properly evaluate the cases according to the coding scheme. For instance, Larsson mentioned his study on mergers and acquisitions, where he employed a minimum requirement of two pages describing the business and human issues. Another aspect is to proper \textbf{sample the cases} to avoid biases like a survey.

\textbf{Conversion from qualitative to quantitative data} should be done by \textbf{multiple raters} using a \textbf{pre-defined coding scheme}. 

\textbf{Analysis of quantitative data} should be based on statistics. 

\subsection{Differences to other research methods}

It is important to compare the case survey with other related, but distinct, research methods common in Software Engineering literature: surveys, multiple-case studies, and systematic mapping studies(SMS)/literature reviews (SLR).

A survey is ``a method to collect and summarize evidence from a large representative sample of the overall population of interest''~\cite{Molleri2016}. It is a research process that is more than a questionnaire, including aspects such as sample size,  population, questionnaire design, response rate, and analysis~\cite{Pfleeger2001} which goal is to generalize~\cite{Stol2018}. Molleri et al.~\cite{Molleri2016} identified 15 articles addressing guidelines for the method in Software Engineering research. In summary, in a survey, a representative sample of a population is probed (generally through a questionnaire) to reach conclusions about the whole population.

A multiple-case study is an inquiry where multiple instances of a phenomenon are investigated. According to Yin~\cite{Yin2003}, a case study is ``an  empirical inquiry that investigates a contemporary phenomenon within its real-life context, especially when the boundaries between phenomenon and context are not clearly evident.'' A multiple-case study is justified when, based on a theoretical framework, the researcher wants to analyze cases where the theory predicts similar results (literal replication) or contrasting (theoretical replication)~\cite{Yin2003}. When comparing to case surveys, Larsson~\cite{Larsson1993} stressed that multiple-case studies``can achieve [cross-case pattern analysis], but the resource-consuming, intensive research they require typically limits case sets to smaller sizes than are needed to benefit from advanced statistical cross-case analysis.'' Besides that, he added that, in contrast to case surveys, multiple-case studies ``are performed by the same person, with the same purpose, perspective, method, and theoretical framework.'' 

Multiple-case studies are mainly based on primary data. This data is ``collected for the specific research problem at hand, using procedures that fit the research problem best''~\cite{Hox2005}. Hox and Boeije~\cite{Hox2005} classified primary data according to two axes: qualitative/quantitative and solicited/spontaneous. Lethbridge et al.~\cite{Lethbridge2005} differentiate three orders of data collection techniques. In the first degree, researchers are in direct contact with subjects and collect data in real-time~\cite{Runeson2009}. Examples are focus groups, survey, and participation observation when the research join the team. In the second degree techniques, researchers collect raw data but without interacting with the subjects. Examples are instrumentation and observations (fly-in-the-wall). Finally, the third degree is characterized by the analysis of available data like tool use logs or documentation analysis.

Regarding systematic reviews or maps, first of all, it is essential to distinguish them. According to Petersen et al.~\cite{Petersen2015}, these methods have different goals and, consequently, research processes. While systematic reviews aim to synthesize evidence, maps focus on structuring a research area. This difference leads to distinct research questions, search process, search strategy requirements, quality evaluation, and results~\cite{Petersen2015}. First, in maps, the research questions focus on discovering research trends while reviews on aggregating evidence for a specific goal. Because of that, the search requirements are less strict for maps and do not have to reach all studies as in a systematic review. Additionally, quality evaluation is not mandatory for maps~\cite{Wohlin2013}. Regarding evidence synthesis, many techniques can be employed, as described by Cruzes and Dyb\aa~\cite{Cruzes2011}. In both cases, though, data is secondary; that is, it was created by other researchers for different purposes~\cite{Hox2005}.

Based on the discussion above, a case survey study is similar to an SLR since its goal is to synthesize evidence rather than structure a research area as in an SMS. Nevertheless, case surveys are constrained to analyze studies describing cases using quantitative techniques on data obtained through a defined coding scheme. Meanwhile, SLRs can, and should, consider studies performed with different research methods, e.g.controlled experiments or surveys, and, as mentioned above, have at disposal different synthesis techniques.

Table~\ref{tab:comparison} presents a summary comparing case survey, multiple-case studies, and literature maps/reviews.

\begin{table*}
	\caption{A comparison between survey, multiple-case study, SLR and case survey}
	\label{tab:comparison}
	\begin{tabular}{m{.07\textwidth}m{.17\textwidth}m{.25\textwidth}m{.19\textwidth}m{.2\textwidth}}
		\toprule
		& Survey & Multiple-case Study & SLR & Case Survey  \\
		\midrule
		Source of Data & Pre-defined instrument, usually a questionnaire. & Several possible data collection methods.               & Literature.   & Literature.               \\
		\hline
		Subjects \newline selection & Sample from the population. & Cases can be selected if they ``typical'', ``critical'', ``revelatory'' or ``unique.'' Cases can be replicated according to theoretical or literal replication. & Study identification method may define venues or publication types. Inclusion and exclusion criteria.  & Based on the research questions and amount of data reported. Should not focus on publication venue.                                \\
		\hline
		Analysis \newline techniques & Usually, statistical analysis. & Quantitative or qualitative analysis. Qualitative are commonly used and could be hypothesis generating, such as constant comparison and cross-case analysis, or hypothesis confirmation, such as triangulation and replication~\cite{Runeson2009}.  & Pre-defined data extraction based on the research questions.   & Coding scheme to convert qualitative into quantitative data. Multiple raters. Statistical analysis of coded data.                                   \\
		\hline
		Goal &  Derive generalizable conclusions regarding the whole population. & Although generally to explore, they can also be descriptive or explanatory.   &  Aggregate conclusions of different studies.      &  Draw new conclusions based on the analysis of quantitative data obtained by applying a common coding scheme to published cases with rich qualitative data.                 \\
		\bottomrule                       
	\end{tabular}
\end{table*}

\section{Research Method}
\label{sec:research_method}

To achieve our goal of exploring how the case survey method has been used in Software Engineering research, we performed a systematic mapping study (SMS) following Petersen et al.~\cite{Petersen2015} guidelines. The authors summarized the steps generally suggested for SMS in four:  identification of the need for the map, studies selection, data extraction and classification, and study validity. 

From the exposed above, there was no study exploring the use of case survey in Software Engineering research. This study aims to fill this void. The first step in this direction is to define research questions (RQs) that should be answered by analyzing the selected primary studies. Such a step has been done in Section~\ref{sec:introduction}.

In the second step, we defined the query string based on PICO (Population, Intervention, Comparison, and Outcomes). Similarly to our methodological reference, i.e.~\cite{Petersen2015} that reviewed SMSs in Software Engineering literature, our query string consisted only on a filter on population, that is, case survey studies performed in Software Engineering research. Based on these two elements, case surveys, and software engineering, we developed the search string considering possible variations of the term case survey. We included the variation ``case study survey'' after an initial query of the databases returned a study that used this term. Therefore, we reviewed the search string and redid the queries on the databases. The final query string is the following:

\begin{center}
	("case survey"  OR  "case-survey"  OR  "case study survey")  AND  "software" 
\end{center}

Petersen et al.~\cite{Petersen2015} recommended performing the search on IEEE, ACM, and two indexing services. Therefore,  we included Scopus and Web of Science. We completed the query against each database filtering all possible fields (including full text). Such restriction was needed because, in an initial evaluation, Papatheocharous et al.~\cite{Papatheocharous2018} did not mention the use of case survey in the abstract but only in the full text.

We created a spreadsheet with all the papers gathered, including title, authors, venue, and year. We removed the duplicates and reached a total of 511 papers. Based on that, we identified the primary studies to be analyzed. In this step, the following inclusion criteria were:

\begin{itemize}
	\item Studies are in the field of software engineering.
	\item Studies employ case survey and claim to do it. Either as a stand-alone method or within a mixed-methods approach.
\end{itemize} 

The exclusion criteria were:

\begin{itemize}
	\item Studies only presenting a research proposal.
	\item Studies were not peer-reviewed.
	\item Summaries or editorials of conferences or proceedings.
\end{itemize}

\begin{figure}[h]
  \centering
  \includegraphics[width=\linewidth]{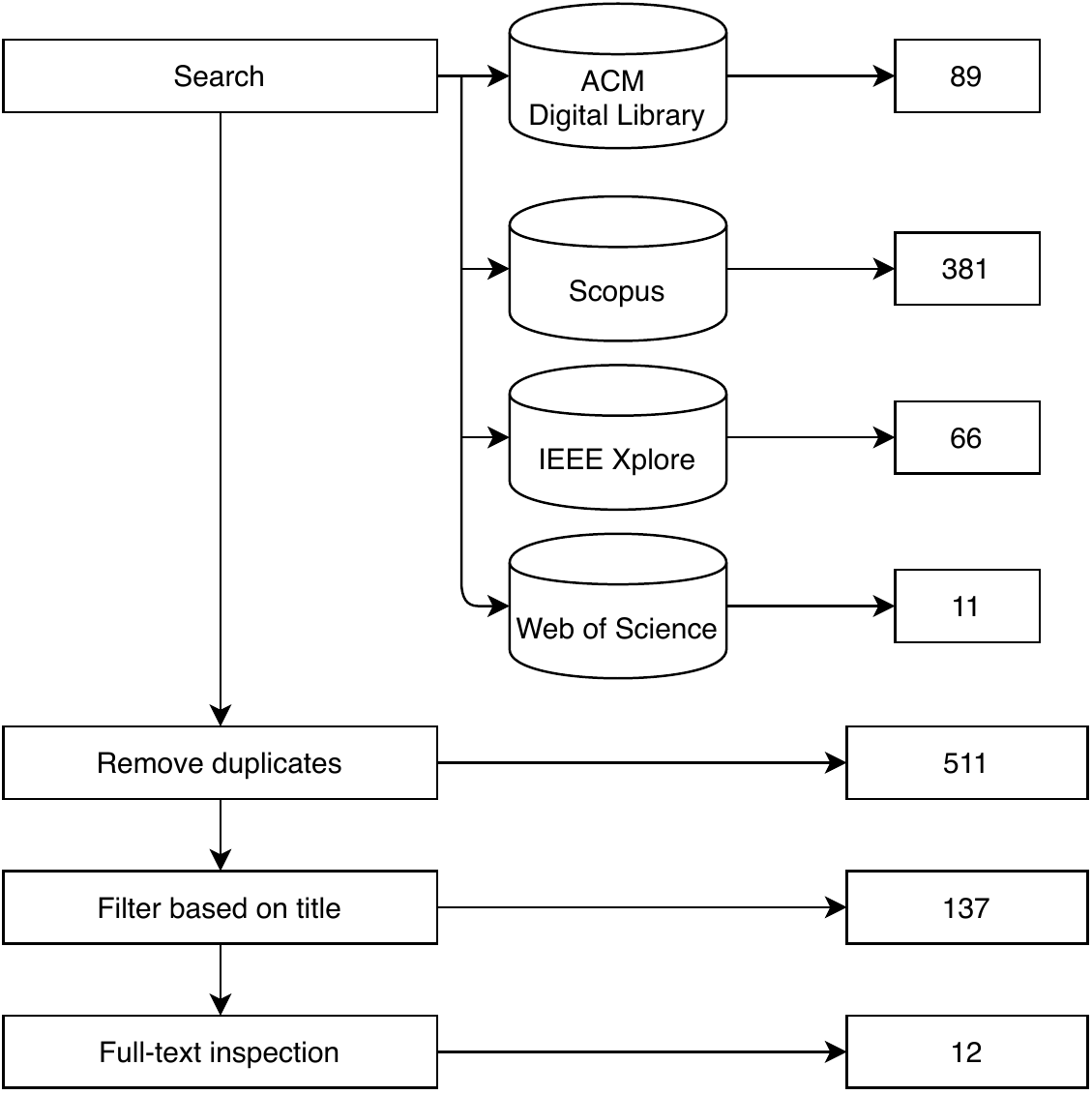}
  \caption{Primary studies selection procedure.}
  \Description{Primary studies selection procedure.}
  \label{fig:sms}
\end{figure}

The first author evaluated the title and venues to remove papers that were obviously not related to software engineering. Examples of excluded papers in this phase are studies focused on the use of software in another field, such as nursing or education. Nevertheless, a conservative approach was taken, and, in case of doubt, the paper was kept to be analyzed in the next step. In this stage, papers that were not research papers, e.g., tables of contents or workshop reports, were also excluded. As a result, 137 papers were included, and their full-text was inspected for further inclusion/exclusion. The final list of primary studies consisted of 12 papers. The intermediate results of this process were made openly available online~\cite{Melegati2020}. Fig.~\ref{fig:sms} summarizes the process.  

In the data extraction and analysis phase, the strategy was discussed and agreed upon between the two authors. Then the first author carried out the task, and the results were inspected by both authors and discussed between them. Besides the title, authors, venue, and year already available from the selection phase, the 12 papers' full-text were analyzed to retrieve their research questions (or goals) and the employed research method. To compare with the canonical case survey, we used the steps previously identified. For case selection, we considered which sources the authors searched for cases and the employed inclusion/exclusion criteria. Regarding data analysis, we observed if and how the authors converted qualitative into quantitative data, including how many raters were used, and, finally, if and how they analyzed the quantitative data.

\section{Results}
\label{sec:results}

This section presents the 12 identified papers and how they answer our research question. Table~\ref{tab:studies} presents a summary of the identified studies, including the research questions they answered.

\begin{table*}
	\caption{The studies that used the case survey methodology.}
	\label{tab:studies}
	\bgroup
	\def\arraystretch{1.3}
	\begin{tabular}{llm{0.25\textwidth}m{0.4\textwidth}}
		\toprule
		Paper & Year & Venue & Research question(s) or objective(s)\\
		\midrule
		Bharosa et al.~\cite{Bharosa2009}  & 2009  &   International Conference on Information Systems for Crisis Response and Management (ISCRAM)  & What are critical information and system quality requirements for multi-agency disaster management and how do information architects deal with these requirements in practice?        \\
		\hline
		El-Masri and Rivard~\cite{El-Masri2012} &  2012  & International Conference on Information Systems (ICIS)  & Distinguish the various patterns in which components of risk interrelate                      \\
		\hline
		Garousi et al.~\cite{Garousi2017} &  2017                     &  International Conference on Evaluation and Assessment in Software Engineering (EASE)   & What are the challenges during data extraction in SLR studies?\newline
		How can effectiveness and efficiency of data extraction be improved to address the challenges identified in the previous RQ?   \\
		\hline
		Bajwa et al.~\cite{Bajwa2017}  &  2017                     &   Empirical Software Engineering   & What are the factors that trigger software startups to pivot? \newline What are the types of pivots software startups undertake?                              \\
		\hline
		Petersen et al.~\cite{Petersen2018}  & 2018                       &  IEEE Transactions on Software Engineering  & How decision making took place when choosing among CSOs for adding components in industrial software-intensive systems?                                 \\
		\hline
		Melegati and Wang~\cite{Melegati2018} &  2018                     &   International Workshop on Software-Intensive Business (IWSiB)  & What defines the innovation of a software startup?                                \\
		\hline
		Papatheocharous et al.~\cite{Papatheocharous2018} &  2018                     &  Information and Software Technology  & Support  the systematic documentation of asset-selection decisions in software development                                  \\
		\hline
		Tripathi et al.~\cite{Tripathi2018}  &  2018                     & Information and Software Technology  & Five RQs regarding sources, elicitation methods, documentation, prioritization and managements, and validation of requirements in software startups.                                   \\
		\hline
		Klotins et al.~\cite{Klotins2018} & 2018  &    International Conference on Software Engineering - Software Engineering in Practice track (ICSE-SEIP) & How do start-ups estimate technical debt? \newline What are precedents of technical debt in start-ups? \newline What outcomes linked to technical debt do start-ups report?               \\
		\hline
		Hyrynsalmi et al.~\cite{Hyrynsalmi2018} & 2018  &  International Conference on e-Business, e-Services and e-Society  (I3E)  &  How computer game start-ups perceive
		and use MVPs in their businesses?                       \\
		\hline
		Klotins et al.~\cite{Klotins2019} & 2019  &   IEEE Transactions on Software Engineering & What patterns pertaining software engineering can be ascertained in start-up companies?                              \\
		\hline
		Garousi et al.~\cite{Garousi2019b} & 2019 & IEEE Software & Present the authors' experiences with industry-academia collaborations. \\
		\bottomrule                       
	\end{tabular}
	\egroup
\end{table*}

In chronological order, the first identified study was performed by Bharosa et al.~\cite{Bharosa2009}. The authors investigated the relevance and assurance of quality requirements for information systems' success during disaster management. To reach their goal, the authors followed a research approach combining literature review, empirical case studies, and semi-structured interviews about four cases (the 9/11 attacks, Hurricane Katrina, the 2005 Asian Tsunami, and the large fire at the Schiphol Detention Complex in the Netherlands). The selection of cases was based on the amount of documentation and complementary disaster source (human vs. natural). More than one published study could have been evaluated for each case. The authors claimed that ``the main objective of our case survey was to identify and describe [information quality] and [software quality] related problems which occurred during the response the disasters.'' 

Comparing how the case survey method was applied to what has discussed above, the number of cases was low (four) and not systematically selected. Although the analysis was based on previously reported data, it consisted of qualitative analysis. Although a coding protocol was used, it was not used to generate quantitative data and, consequently, no quantitative analysis was performed. 

El-Masri and Rivard~\cite{El-Masri2012} investigated design principles for software project risk management systems. To achieve this goal, the authors analyzed 60 software projects and observed patterns ``of [the] interplay among risk components.'' The cases investigated were identified by searching the literature for papers describing software projects. The results consisted of four patterns: ``the multiplicative effect of project traits, the sequentiality effect of undesirable events, the presence of a third variable, and the tradeoff when implementing risk management practices.''

Regarding the application of case survey, the authors performed a search for previously published papers, but they argued to have stopped when ``theoretical saturation'' was reached. Multiple raters performed coding based on a previously defined model. Nevertheless, the analysis is predominantly qualitative by grouping codes according to relationships among them.  

Garousi et al.~\cite{Garousi2017} analyzed data extraction techniques used in systematic reviews in software engineering. The authors claimed to have used case survey and thematic synthesis to synthesize the data. The case survey was mainly used ``to collect data (challenges in data extraction) about each case. '' As the data source, they used 16 systematic literature reviews the authors have been involved and a review of challenges and guidelines existent in the literature. 

Regarding the research method application, the considered cases consisted of the authors' own previously published papers. Therefore, no search was performed. Besides that, the analysis consisted of thematic synthesis based on ``open and axial coding.''

Bajwa et al.~\cite{Bajwa2017} studied the pivots, a strategic change of a business concept, product, or the different elements of a business model, software startups make during their history. They also analyzed the triggers that led startups to these changes. The study comprehended an analysis of 49 pivoted software startups in which data was found online through a systematic search on Google. As a result, the authors identified ten pivot types and 14 triggering factors.

Although the number of ``cases'' surveyed is large, they were not published case studies but, instead, statements from founders or the press about pivots. The authors also employed qualitative analysis to identify patterns on the data that represented different pivot types or triggers.

Petersen et al.~\cite{Petersen2018} explored how decision making took place in the industry to choose among component source options (CSOs). The authors studied 22 industrial cases through a pre-defined ``data extraction scheme'' consisting of interviews that participated in the projects. The cases were conveniently sampled through the authors' contacts and projects in which they were involved. The results showed that decisions ``are deterministic and based on optimization approaches.'' One interesting aspect of this paper is the number of authors, ten, that could explain how they managed to study such a large number of cases.

Compared to the case survey method, the study was not based on previously published cases but, instead, on a convenience sample. The authors provided a coding scheme to researchers report data on these cases. Based on that, the study presented a statistical analysis of the data.

Melegati and Wang~\cite{Melegati2018} investigated how software startup researchers understood the concept of innovation used to describe these companies. To reach this goal, the authors analyzed 27 published papers from the period 2013-17, identifying 18 software products or services. Then, they classified these solutions based on which type of discontinuity the innovation represented: technological, market-related, or both. The authors concluded that research on software startups had not differentiated companies based on the innovation they propose. 

Regarding the method, the authors based their list of published papers in a search previously performed in another systematic mapping study, and they considered the cases according to the amount of data available. Consequently, the number of cases (18) was low. The conversion was related to only two categorical variables (presence or not of technological and marketing innovation), and the statistics were based on frequency. There is no mention to multiple raters and, consequently, inter-rate agreement statistics.

Papatheocharous et al.~\cite{Papatheocharous2018} used the case survey methodology to evaluate a proposed taxonomy, called GRADE, for documentation supporting decision-making in the development of software-intensive systems. Based on requirements identified through a previous literature survey on systematic decision-making, the authors developed the taxonomy following a defined approach. The result consisted of a ``structured way of characterizing decisions for software-intensive systems through the lenses of five top-level elements, namely Goals (G), Roles (R), Assets (A), Decision methods and criteria (D), and environment."  The artifact was evaluated in two phases: first with four decision cases from two companies and 29 industrial decisions, for a total of 33 cases. Five researchers used the taxonomy to evaluate the industrial decisions, and their agreement was analyzed using the Jaccard index.

A particular aspect of applying the case survey method in this study was its use for validating an artifact proposed in the same study. Therefore, it is clear that it was not based on previously published studies but a convenience sample. Similar to Petersen et al.~\cite{Petersen2018}, the authors used a scheme to code the cases they were analyzing and performed some quantitative analysis on it. It is interesting, though, that five researchers independently coded the decisions.

The next four papers presented studies based on the same data so that we will discuss them together. Data consisted of answers of software startup members to a  questionnaire that captured ``all relevant data about a case of interest''~\cite{Klotins2017}. The data collection process was described by a workshop paper (i.e.,~\cite{Klotins2017}), excluded from our mapping study because it was only a research proposal. In this paper, the author cited Larsson~\cite{Larsson1993}, saying that the method was initially proposed for a secondary study. Then, he suggested an adaption ``for use in primary studies and to collect data directly from people involved in the cases. '' The questionnaire contained 89 questions either open or multiple-choice.

The first paper to present this questionnaire results was a mixed-method study on requirements engineering in software startups by Tripathi et al.~\cite{Tripathi2018}. Besides the case survey, the authors also conducted a multi-vocal literature review. In this study, the authors acknowledged 80 responses to the questionnaire. The results comprehended a description of how these companies perform requirements activities such as elicitation and documentation.

Then, the following studies considered 86 responses to the questionnaire. First, Klotins et al.~\cite{Klotins2018} explored the concept of technical debt in software startups. The results identified dimensions, precedents, and outcomes of technical debt. In another paper~\cite{Klotins2019}, the same authors proposed a progression model for software startups consisted of four stages (inception, stabilization, growth, and maturity). They identified goals, challenges, and practices common to these companies and mapped them into the stages. Hyrynsalmi et al.~\cite{Hyrynsalmi2018} used a subset of nine answers to understand how game developing startups are applying the Minimum Viable Product (MVP) concept. The results showed that the views on the topic are divergent and practical guidelines were still lacking.

Regarding the case survey method, as mentioned before, data is primary, that is, not previously published. Instead of a coding scheme and reader-analysts, the researchers applied an instrument (a questionnaire) to subjects. Given the questionnaire depth, the authors considered each answer a case. Therefore, there were no multiple raters for each case. Among the papers, the authors analyzed the data differently but mainly qualitatively or based on frequency statistics.

Garousi et al.~\cite{Garousi2019b} explored industry-academia collaboration in Software Engineering researchers. To achieve their goal, the authors synthesized evidence from 26 international projects in which they were involved. The results comprised a summary of the experience and lessons learned.

Similarly to the above-described study performed by the same group of researchers~\cite{Garousi2017}, the study was based on previously published studies by the authors. No search for other cases was performed, and analysis consisted of qualitative techniques. 

Tables~\ref{tab:case_selection} and~\ref{tab:case_survey_on_papers} summarize how the studies employed the case survey methodology according to the aspects described in the seminal methodological papers.

\begin{table*}
	\centering
	\caption{The case selection strategies of the reviewed studies.}
	\label{tab:case_selection}
	\bgroup
	\def\arraystretch{1.3}
	\begin{tabular}{lrm{0.33\textwidth}m{0.32\textwidth}}
		\toprule
		Paper & \parbox{0.07\textwidth}{Number\\ of cases} & Source & Inclusion and exclusion criteria  \\
		\midrule
		Bharosa et al.~\cite{Bharosa2009}  & 4 & Pre-defined cases. Papers in the literature about these cases. & The cases were selected based on the amount of documentation available and complementary regarding disaster source.             \\
		\hline
		El-Masri and Rivard~\cite{El-Masri2012}  & 60 & Papers in the literature describing software projects. & The study had a software project as a unit of analysis and the article provided enough information about the project context and execution. Stopped with theoretical saturation. \\
		\hline                   
		Garousi et al.~\cite{Garousi2017}  &  16 & SLRs in which the authors were involved. & None.                       \\
		\hline
		Bajwa et al.~\cite{Bajwa2017}    &  49 & Web pages found using Google search. & Several criteria including that the website depicts a software startup pivot.  \\
		\hline
		Petersen et al.~\cite{Petersen2018}    & 22 & Sampled from authors' contacts and industrial projects. Data came from recollections of researchers that interacted previously with the cases and interviews with practitioners. & None.                           \\
		\hline
		Melegati and Wang~\cite{Melegati2018} & 18 & Papers in the literature describing software startups. & Enough information about the product.                            \\
		\hline
		Papatheocharous et al.~\cite{Papatheocharous2018} & 33 & The authors investigated four cases from two companies and 29 from ``own experiences and interviews of past decision-making.'' & None.    \\
		\hline
		Tripathi et al.~\cite{Tripathi2018}      &   80 & Convenience sampling. Data consisted of answers to a questionnaire. Each answer was considered a case.  & Complete answers to the questionnaire.                        \\
		\hline
		Klotins et al.~\cite{Klotins2018}  &  86 & Same as Tripathi et al.~\cite{Tripathi2018}   &   None.                       \\
		\hline
		Hyrynsalmi et al.~\cite{Hyrynsalmi2018}  &   9  & Same as Tripathi et al.~\cite{Tripathi2018} & Startups that developed games.               \\
		\hline
		Klotins et al.~\cite{Klotins2019} &  86 & Same as Tripathi et al.~\cite{Tripathi2018} &  None.                      \\
		\hline
		Garousi et al.~\cite{Garousi2019b}  & 26 & Projects where the authors were involved. & None. \\
		\bottomrule                       
	\end{tabular}
	\egroup
\end{table*}

\begin{table*}
	\centering
	\caption{The data analysis strategies of the reviewed studies.}
	\label{tab:case_survey_on_papers}
	\bgroup
	\def\arraystretch{1.3}
	\begin{tabular}{lm{0.24\textwidth}m{0.24\textwidth}m{0.24\textwidth}}
		\toprule
		Paper & Conversion of qualitative into quantitative data & Raters & Quantitative data analysis  \\
		\midrule
		Bharosa et al.~\cite{Bharosa2009}  & No. & Not applicable.  & No. \newline Qualitative content analysis using a coding protocol.   \\
		\hline
		El-Masri and Rivard~\cite{El-Masri2012}  & No. \newline Existing theoretical model as coding scheme. & Information not available. & No. \newline Relationships among codes were judged based on qualitative techniques like matrices.  \\
		\hline
		Garousi et al.~\cite{Garousi2017}  & No. \newline Case survey was used to collect data regarding challenges through qualitative analysis. & Information not available.  & No. \newline Thematic synthesis was used and challenges were grouped.                \\
		\hline
		Bajwa et al.~\cite{Bajwa2017}    & Open coding (frequency of the emergent codes).  & Performed by the two authors separately and then compared. &  Frequency analysis.                   \\
		\hline
		Petersen et al.~\cite{Petersen2018}    & A data extraction form was used on data provided by researchers and from the interviews with practitioners. Open coding was used on some answers.  &  Initially coded by the first author and then reviewed by two others.   & Vote counting and odds ratio.                      \\
		\hline
		Melegati and Wang~\cite{Melegati2018}  &  One dimension: type of discontinuity. & Not mentioned in the paper.  & Frequency analysis in a matrix.                       \\
		\hline
		Papatheocharous et al.~\cite{Papatheocharous2018} &  The cases data was used to evaluate the proposed taxonomy. & Independently  by two researches from a group of five. & Jaccard index was used to compare the taxonomy use among the researchers.                         \\
		\hline
		Tripathi et al.~\cite{Tripathi2018}   & Integrated approach (inductive and deductive). & Not mentioned in the paper. &  Descriptive statistics, bivariate correlations, multiple-response cross-tabulations. \newline Qualitative analysis was also employed with an integrated approach, i.e. deductive and inductive methods.  \\
		\hline
		Klotins et al.~\cite{Klotins2018}  & Respondents estimates were converted to an ordinal scale.  & Not mentioned in the paper.  & Descriptive statistics and association tests (Chi-square and Cramer's V tests).                         \\
		\hline
		Hyrynsalmi et al.~\cite{Hyrynsalmi2018}  & No.  & Not applicable. & No. \newline Qualitative analysis of answers.\\
		\hline                        
		Klotins et al.~\cite{Klotins2019} & Some answers to the questionnaire were already quantitative, others were coded using open coding.  & Initially by the first authors. Then jointly evaluated by the first and second authors. & Descriptive statistics, contingency tables, and association tests. \newline Qualitative analysis was also used (open coding).                       \\
		\hline
		Garousi et al.~\cite{Garousi2019b}  & No. & Not applicable. & No. \newline Meta-analysis to synthesize the authors' experiences.  \\
		\bottomrule                       
	\end{tabular}
	\egroup
\end{table*}

The differences observed in the application of the research method could be a consequence of different methodological references. Nevertheless, the most common reference (used on nine of the analyzed studies) was Larsson~\cite{Larsson1993}. 

Garousi et al.~\cite{Garousi2017} referenced the book by Runeson et al.~\cite{Runeson2012} on case study research. In this book, the authors referenced Larsson~\cite{Larsson1993} and summarized ``the case survey method aggregates existing case study research by applying a set of structured and tightly defined questions and answers, to each primary study.'' 

Bajwa et al.~\cite{Bajwa2017} referenced Cruzes et al.~\cite{Cruzes2015} that also cited Larsson~\cite{Larsson1993}. They described the method as  a ``formal process for systematically coding relevant data from a large number of case studies for quantitative analysis, allowing statistical comparisons across studies.'' The authors also compared it to other synthesis methods such as thematic or narrative synthesis.

Another reference from the same group of authors is Cruzes and Dyb\aa~\cite{Cruzes2010}, cited by Garousi et al.~\cite{Garousi2019b}. Here, they said that ``each primary study [is] treated as a `case'. Study findings and attributes [are] extracted using closed-form questions, for reliability. Survey analysis methods [are] used on extracted data.'' The authors cite Yin and Herald~\cite{Yin1975}.

In summary, except by one study, all primary studies relied, directly or indirectly, on Larsson~\cite{Larsson1993} as the methodological reference. It is then interesting to observe how the actual implementations diverged from what was initially described. Table~\ref{tab:references} displays the references for all the analyzed papers.

\begin{table}
	\caption{Methodological references on case survey used in the reviewed studies.}
	\label{tab:references}
	\begin{tabular}{ll}
		\toprule
		Methodological reference & Papers \\
		\midrule
		\multirow{9}{*}{Larsson~\cite{Larsson1993}}& Bharosa et al.~\cite{Bharosa2009}  \\
		& El-Masri and Rivard~\cite{El-Masri2012}   \\
		& Petersen et al.~\cite{Petersen2018}    \\
		& Melegati and Wang~\cite{Melegati2018} \\
		& Papatheocharous et al.~\cite{Papatheocharous2018} \\
		& Tripathi et al.~\cite{Tripathi2018}   \\
		& Klotins et al.~\cite{Klotins2018}    \\
		& Hyrynsalmi et al.~\cite{Hyrynsalmi2018}   \\
		& Klotins et al.~\cite{Klotins2019} \\
		\hline
		Runeson et al.~\cite{Runeson2012}  & Garousi et al.~\cite{Garousi2017}  \\
		\hline
		Cruzes et al.~\cite{Cruzes2015}  & Bajwa et al.~\cite{Bajwa2017}  \\
		\hline
		Cruzes and Dyb\aa~\cite{Cruzes2010} & Garousi et al.~\cite{Garousi2019b}  \\
		\bottomrule                       
	\end{tabular}
\end{table}

\section{Discussion}
\label{sec:discussion}

The SMS results indicated that the case survey method had been rarely used in Software Engineering research. Nevertheless, its use has increased in the last 3-4 years since 10 of the 12 primary studies were published in this period. Like what happened to other ``imported'' research methods, we could observe a lack of adherence to what seminal methodological papers previously prescribed. 

The most common type of research approach described as a case survey was a method based on primary, but limited, data from a large number of instances of the phenomenon under study that the authors called ``cases.'' Data consisted of answers to a questionnaire (in the case of the startup studies that used the same data to study different aspects of these companies ~\cite{Tripathi2018,Klotins2018,Hyrynsalmi2018,Klotins2019}) or online resources, i.e., blog posts, to investigate software startup pivots~\cite{Bajwa2017}.

Using the term case survey to describe an analysis of a large set of phenomenon instances through the application of a questionnaire or an analysis of blog posts or other online available resources is a strong extrapolation. Such a claim implicitly refers to each analyzed questionnaire answer/resource as a case study. But, the depth and breadth of the data about that case, generally a questionnaire answer or a blog post from only one person, weakens this claim. First, the case study is a research method and, as such, requires a systematic use to provide reliable results. This need is corroborated by the numerous guides proposed in the literature, either in social sciences (e.g.,~\cite{Yin2003}) or in Software Engineering (e.g.,~\cite{Wohlin2012} and~\cite{Runeson2009}). Although Larsson~\cite{Larsson1993} gave examples of the use of non-peer-reviewed cases, such as articles in the Fortune magazine, his examples consisted only of descriptions performed by people, not from the case (like the Fortune writers). The author stressed the need for data deeply describing the case (for instance, in his study on mergers and acquisitions, he applied the criteria of at least two pages describing the company). He also made a clear distinction between case and multiple-case studies, as mentioned earlier.

Second, using only one source to describe a case jeopardizes one key element for case studies: triangulation. Yin~\cite{Yin2003}, in his seminal book about the methodology,  defines it, among other aspects, as an inquiry that ``relies on multiple sources of evidence, with data needing to converge in a triangulating fashion.'' Runeson and H\"ost~\cite{Runeson2009} also mentioned: ``it is important to use several data sources in a case study in order to limit the effects of one interpretation of one single data source.'' Although convergence may not be the only result of triangulation, as it can lead to inconsistency or contradiction~\cite{Mathison1988}, it helps improve the study results. For instance, one of the elements of validity for case studies is internal validity that concerns causal relationships and a possible erroneous inference of a factor determining a consequence when, in reality, another factor not considered in the analysis is the determinant~\cite{Runeson2009}. 

Once the instances observed are not cases, these studies are not secondary. That is, the data collected is primary, and this is another argument against naming this empirical setting as a case survey, generally seen as a review method~\cite{Jurisch2013}.

To strengthen our argument, you will use Stol and Fitzgerald's ABC framework~\cite{Stol2018}. Adapting from social sciences, the authors proposed this framework consisted of two dimensions: the research obtrusiveness level and results generalizability. Based on that, they argued that it is impossible to optimize a study to reach generalizability over actors (A), precise measurement of their behavior (B) in a realistic context (C). The authors positioned different empirical strategies according to these characteristics. In this sense, sample studies, such as surveys, represent less obtrusive research that does not focus on specific contextual details but a universal context and system, to reach greater generalizability over actors (A). In this sense, the case survey method, as described by Larsson~\cite{Larsson1993}, overcomes this limitation by being a secondary study where the several case studies achieved realism of context (C) and a large number of these studies reached some generalizability over actors (A).

Most of the studies analyzed in this SMS did not focus on the realism of the context. Such an element is not achievable neither with a detailed questionnaire nor a blog post. As mentioned before, case studies are based on data triangulation to achieve this aspect. These studies focus' is on generalizability through analyzing several instances of the phenomenon. Other studies performed something similar but were restricted to published papers where the authors were involved. In this category, we could include \cite{Garousi2017} and \cite{Garousi2019b}.

Nevertheless, these studies are similar to what Jansen~\cite{Jansen2010} called a qualitative survey distinguishing it from qualitative surveys. While in the latter, one counts the frequencies of categories or values, in the first, one searches for ``empirical diversity in the property of members.'' Such a goal is clear on research questions that aim to identify different aspects of the software startup population.

Another understanding of the case survey term focused on an analysis of a large number of cases. In this category, we could mention ~\cite{Petersen2018} and ~\cite{Papatheocharous2018}. In both studies, the authors analyzed a large number of cases (22 and 33). Nevertheless, they present some differences. Petersen et al.~\cite{Petersen2018} collected data from multiple sources for each case and performed a quantitative comparison of the cases. It diverged from a canonical case survey in the sense of analyzing previous studies but, instead, executing themselves.
On the other hand, Papatheocharous et al.~\cite{Papatheocharous2018} claimed to use case survey to validate a proposed artifact. The data used was deeper only in four cases; the other 29 were based on limited data in a similar way to Petersen et al. Both studies are in the boundary between multiple-case studies and case surveys.

Such lack of adherence between a research method and how Software Engineering researchers employ it has already been found before. In their study on the use of grounded theory in Software Engineering research, Stol et al.~\cite{Stol2016} observed several occurrences of ``method slurring''. That is, several articles ``claim to use grounded theory, yet do not embrace its core characteristics.'' The authors also suggested five reasons that could explain this phenomenon: to confer legitimacy, to avoid detailed and exhaustive literature review and initial conceptualization, for simplicity, lack of understanding of the method, or per referee's suggestion.

This lack of adherence to the described methods is detrimental to Software Engineering research for several reasons. First, it compromises the quality of a study. Since the underlying epistemological assumptions of a defined research method have already been discussed, the findings from a study consistently applying such a method could be trusted. When the use deviates from the canonical approach, without proper reflection on the consequences, the validity of the research findings may be put into question. Second, it compromises the work of reviewers. A defined research method represents a checklist for reviewers while evaluating a manuscript. The divergence from it makes the evaluation process more complex.

\subsection{Threats to validity}

In their SMS guidelines, Petersen et al.~\cite{Petersen2015} used five aspects in their discussion about threats to validity: descriptive validity, theoretical validity, generalizability, interpretive validity, and repeatability.

Descriptive validity regards how accurately and objectively the observations describe the reality. To improve this aspect, we used a defined set of elements based on the definition of the case survey method that should be observed in the primary studies. 

Theoretical validity concerns the correct identification of the aspects that were the goal of the study. Threats to this validity element could happen in the study selection and data extraction steps. First, published studies that should be included could not have been found. To mitigate this threat, we had an initial set of studies that were used to evaluate the query string. Besides that, the query string was updated after a new term, case study survey, was found. In this regard, another issue could have been the exclusion of a paper when it should not be. To avoid this issue, we used a conservative approach when filtering by title. A possible criticism is that we may not have considered studies that could be classified as case surveys but have not labeled them as so. Since the research goal was to understand how the method is being understood and employed in Software Engineering research, we think not considering these papers was correct. Another threat in this category occurs during data extraction. Like Petersen et al.~\cite{Petersen2015}, to mitigate this threat, one author extracted the data while another reviewed.

Regarding generalizability, the number of identified studies is low, but we performed a selection procedure that aimed to gather the largest number of studies. Interpretative data regards the coherence between data and analysis conclusions. A common threat in this aspect is the researcher bias that we mitigated using a data extraction scheme based on previous methodological papers on the case survey method. Finally, we achieved repeatability by describing all steps performed.

\section{Conclusions}
\label{sec:conclusions}

In the last 20 years, the human aspects of software engineering led researchers to use methods generally employed in social sciences such as case study and ground theory. More recently, another imported method started to be applied: case survey. This paper's goal was to investigate how Software Engineering researchers have employed this method so far. To achieve it, we performed a systematic mapping study to identify the Software Engineering studies using the method. We identified 12 peer-reviewed publications. Our results indicated that, although the number of studies employing the method rose in the last 3-4 years, the adherence to the method as described in the seminal methodological papers has been low. The most common approach labeled as case survey is actually based on primary data collected through questionnaires or online resources such as blog posts. The other authors used the term case survey to describe multiple-case studies with a large number of cases.

With the results of this study, we aim to raise the awareness of a lack of adherence from the way case survey is employed in the Software Engineering literature to the seminal methodological papers. Our study contributes to a better understanding of case survey as an increasingly used research approach in the field. A good understanding of this approach, which is less familiar to Software Engineering researchers, is beneficial for various groups of the community. To authors, it can help them to communicate better the research method employed, and follow the guidelines to reduce threats to validity and increases reliability. To reviewers and readers, it can help them better assess the quality of the studies employing the case survey method. In our future work, we aim to adapt the case survey guidelines to Software Engineering literature, including the challenges researchers may face.

\bibliographystyle{ACM-Reference-Format}
\bibliography{paper}

\end{document}